\documentclass[prd,twocolumn,nofootinbib,notitlepage,10pt,aps]{revtex4-1}
\usepackage[utf8]{inputenc}
\usepackage{cancel}
\usepackage{xcolor}
\usepackage{latexsym}
\usepackage{amsmath}
\usepackage{amssymb}
\usepackage{bbm}
\usepackage{tensor}
\usepackage{natbib}
\usepackage{placeins}

\usepackage{hyperref}
\hypersetup{
    colorlinks,
    linkcolor={violet},
    citecolor={teal},
    urlcolor={teal}
}
\usepackage{orcidlink}

\makeatletter
\newcommand{\subseconly}[1]{\expandafter\@secondoftwo\csname r@#1\endcsname}
\makeatother

\usepackage{ulem}
\usepackage{pdfsync}
\usepackage{epsfig}
\usepackage{epstopdf}

\usepackage{subcaption}
\usepackage{color}
\usepackage{comment}
\usepackage{slashed}
\usepackage{physics}

\def\beq{\begin{equation}}
\def\eeq{\end{equation}}
\def\baq{\begin{eqnarray}}
\def\eaq{\end{eqnarray}}

\newcommand{\be}{\begin{equation}} 
\newcommand{\ee}{\end{equation}}
\newcommand{\bea}{\begin{equation} \begin{aligned}}
\newcommand{\eea}{\end{aligned} \end{equation}}

\begin{document}

\title{Inflationary assessment of $F(\mathcal{R},\tilde{\mathcal{R}})$ Einstein-Cartan models}

\author{Theodoros Katsoulas\orcidlink{0000-0003-4103-7937}}
\email{th.katsoulas@uoi.gr}
\affiliation{Physics Department, University of Ioannina, 45110, Ioannina, Greece}

\author{Kyriakos Tamvakis\orcidlink{0009-0007-7953-9816}}
\email{tamvakis@uoi.gr}
\affiliation{Physics Department, University of Ioannina, 45110, Ioannina, Greece}

\begin{abstract}
In the framework of $F({\cal{R}},\,\tilde{\cal{R}})$ Einstein-Cartan gravity with an action depending both of the Ricci scalar and the so-called Holst-invariant curvature we consider models that include cubic terms of the latter in the action and study their inflationary behavior. These terms can have a considerable effect either positive or negative in relation to the agreement with present observational data, depending on parameters. In parameter regions where the quadratic models fail to produce results consistent with observational data, the presence of these additional cubic terms can lead to compatible predictions.

\end{abstract}

\maketitle

%%%%%%%%%%%%%%%%%%%%%%%%%%%%%%%%%%%%%%%%%%%%%%%%%%%%%%%%%%%%%%%%%%%%%%%%%%%%%%%%%%%%%%%%%%%%%%%%%%%%
%

\section{Introduction}
\label{Introduction}
	Current cosmological investigations are carried out in the framework of general relativity (GR) as a theory of gravity and the Standard Model (SM) as a theory of matter interactions, while it is assumed that the quantum character of the former could be ignored below the Planck scale in contrast to the full quantum character of the latter. Nevertheless, the quantum interactions of gravitating matter fields are expected to introduce modifications to the standard GR action which can have phenomenological implications. Such modifications are nonminimal couplings of matter fields to curvature or higher order curvature terms~\cite{Davies:1977ze,Birrell:1982ix}. This has motivated the study of more general formulations of the theory of gravitation such as the so-called {\textit{metric-affine }} gravity in which the metric and the connection are independent variables~\cite{Palatini1919, Hehl:1994ue}. In this framework in contrast to standard GR the {\textit{torsion}} and {\textit{nonmetricity}} are in general nonvanishing. A more specific framework is that of the {\textit{Einstein-Cartan}} theory of gravity~\cite{Cartan:1923zea, Cartan:1924yea} where the nonmetricity vanishes while the torsion is present. 
	
	A central paradigm in our present  understanding of the early Universe is provided by cosmological {\textit{inflation}}~\cite{Kazanas:1980tx, Sato:1980yn, Guth:1980zm, Linde:1981mu}, which provides a mechanism for the generation and formation of the observed large- scale structure. Inflation is modeled in terms of a scalar field (the {\textit{inflaton}}), the vacuum energy of which drives the expansion, while its primordial quantum fluctuations~\cite{Starobinsky:1979ty, Mukhanov:1981xt, Hawking:1982cz, Starobinsky:1982ee, Guth:1982ec, Bardeen:1983qw} lead to the inhomogeneities detected in the cosmic microwave background (CMB). The inflaton can be either a fundamental scalar field (e.g. the SM Higgs field~\cite{Bezrukov:2007ep}) or arise as an extra scalar component of gravity itself (e.g. the Starobinsky model~\cite{Starobinsky:1980te}). Another example of the latter case is provided by the framework of Einstein-Cartan gravity, where a pseudoscalar associated with the so-called {\textit{Holst invariant}}~\cite{Hojman:1980kv, Nelson:1980ph, BarberoG:1994eia, Immirzi:1996di, Holst:1995pc} curvature term plays that role~\cite{Pradisi:2022nmh, Gialamas:2022xtt, Salvio:2025izr, Iosifidis:2025mcb, Karananas:2025xcv}.

    In the present article we consider the Einstein-Cartan framework of gravity and focus on gravitational actions of the form ${\cal{S}}=\frac{1}{2}\int\,{\rm d}^4x\,\sqrt{-g}\,F({\cal{R}},\,\tilde{\cal{R}})$, where ${\cal{R}}$ is the Ricci scalar curvature and $\tilde{\cal{R}}$ is the Holst invariant. It has been established that actions at most quadratic in these invariant can lead to models of acceptable inflationary behavior in comfortable agreement with observational data~\cite{ Pradisi:2022nmh, Gialamas:2022xtt}~\cite{Meng:2004yf, Borunda:2008kf, Bombacigno:2018tyw, Enckell:2018hmo, Antoniadis:2018ywb, Antoniadis:2018yfq, Tenkanen:2019jiq, Edery:2019txq, Giovannini:2019mgk, Edery:2019bsh, Gialamas:2019nly, Lloyd-Stubbs:2020pvx, Antoniadis:2020dfq,  Ghilencea:2020piz, Das:2020kff, Gialamas:2020snr, Ghilencea:2020rxc, Iosifidis:2020dck, Bekov:2020dww, Dimopoulos:2020pas,Karam:2021sno, Lykkas:2021vax, Gialamas:2021enw, Antoniadis:2021axu,  Gialamas:2021rpr, AlHallak:2021hwb, Karananas:2021gco, Dimopoulos:2022tvn, Dimopoulos:2022rdp, Durrer:2022emo, Salvio:2022suk, Antoniadis:2022cqh, Gialamas:2022gxv,  Lahanas:2022mng,Iosifidis:2022xvp,Gialamas:2023aim, Gialamas:2023flv,SanchezLopez:2023ixx,DiMarco:2023ncs,Gialamas:2023emn, Gomes:2023xzk, Hu:2023yjn, Gialamas:2024iyu, Gialamas:2024jeb, Racioppi:2024pno, Gialamas:2024uar, Katsoulas:2025mcu, Karananas:2025fas, Karananas:2025qsm}. In this article we investigate whether the presence of terms of higher than quadratic powers of the curvature invariants can still provide us with models of acceptable inflationary behavior~\cite{BERKIN1990348,Huang:2013hsb,Asaka:2015vza,Bhattacharya:2017uwi,Rodrigues-da-Silva:2021jab,Ivanov:2021chn,Wang:2023hsb,Kim:2025dyi,Toyama:2024ugg,Ketov:2025nkr, Addazi:2025qra, Gialamas:2025ofz}. We consider and analyze two such models featuring cubic curvature terms, deformations of inflationary successful examples. We find that the cubic modifications can have a considerable effect in relation to the agreement with observational data depending on the parameter values. In parameter regions where the quadratic models fail to produce results consistent with observations the modified models can lead to acceptable predictions. Although this increases the number of successful candidate inflation models, this might not be the case in the near future when the precision of incoming observational data increases, as is the characteristic case of the until recently robust Starobinsky model which had to be modified in the light of present ACT data~\cite{Salvio:2025izr}~\cite{Kallosh:2025rni, Gialamas:2025kef, Gialamas:2025ofz, Byrnes:2025kit, Addazi:2025qra, Frolovsky:2025iao, Pallis:2025nrv, Wolf:2025ecy, Ahmed:2025rrg, Mohammadi:2025gbu, Yogesh:2025wak, Pallis:2025gii, Ketov:2025cqg, Ellis:2025ieh, Linde:2025pvj, Pallis:2025vxo, Ellis:2025bzi, Ellis:2025zrf}. The article is organized as follows: in Sec.~\ref{Framework} we outline the framework of Einstein-Cartan gravity and derive the Einstein frame action of a general $F({\cal{R}},\,\tilde{\cal{R}})$ model in its equivalent metric form, having integrated out the torsion. In Sec.~\ref{Models} we consider two particular models that include cubic terms of the Holst invariant, one of which is invariant under Weyl transformations, and work out their actions. In Sec.~\ref{Inflation} we proceed to study in detail their inflationary predictions. Finally, in Sec.~\ref{Conclusions} we state briefly our conclusions.

\section{Framework}
\label{Framework}
The Einstein-Cartan formulation of gravity, based on promoting the Poincaré symmetry into a local symmetry, features the {\textit{tetrad}} and the {\textit{spin connection}} as gauge fields with the corresponding field strengths being the {\textit{torsion}} and the {\textit{curvature}}. The former are represented by the metric $g_{\mu\nu}$ and the affine-connection $\tensor{\Gamma}{^\rho_\mu_\nu}$, the latter giving the {\textit{torsion tensor}} and the {\textit{affine curvature}} as
\be\begin{split}
&\tensor{\mathcal{R}}{^\rho_\mu_\nu_\sigma}=\partial_{ \mu}\tensor{\Gamma}{^\rho_\nu_\sigma}-\partial_{ \nu}\tensor{\Gamma}{^\rho_\mu_\sigma}+\tensor{\Gamma}{^\rho_\mu_\lambda}\tensor{\Gamma}{^\lambda_\nu_\sigma}-\tensor{\Gamma}{^\rho_\nu_\lambda}\tensor{\Gamma}{^\lambda_\mu_\sigma}\\\\&
\tensor{T}{^\rho_\mu_\nu}\,=\,\tensor{\Gamma}{^\rho_\mu_\nu}-\tensor{\Gamma}{^\rho_\nu_\mu}.
\end{split}
\ee
Two scalar quantities, linear in the curvature, can be formed, namely, the {\textit{Ricci}} scalar ${\cal{R}}=g^{\mu\nu}{\cal{R}}^{\rho}_{\,\,\mu\rho\nu}$ and the {\textit{Holst invariant}}, as follows:
\be \tilde{\cal{R}}=\epsilon^{\mu\nu\rho\sigma}{\cal{R}}_{\mu\nu\rho\sigma}\ee
The torsion can be decomposed as
\be T_{\mu\nu\rho}=\frac{1}{3}\left(g_{\mu\rho}T_{\nu}-g_{\mu\nu}T_{\rho}\right)+\frac{1}{6}\epsilon_{\mu\nu\rho\sigma}\hat{T}^{\sigma}+\tau_{\mu\nu\rho}\ee 
in terms of a {\textit{torsion vector}} $T_{\mu}=\tensor{T}{^\nu_\mu_\nu}$, an {\textit{axial torsion vector}} $\hat{T}^{\mu}=\epsilon^{\mu\nu\rho\sigma}T_{\nu\rho\sigma}$ and a {\textit{tensorial part}} $\tau_{\mu\nu\rho}$, defined by the constraints $\tensor{\tau}{^\nu_\mu_\nu}=\tensor{\tau}{^\nu_\nu_\mu}=\epsilon^{\mu\nu\rho\sigma}\tau_{\nu\rho\sigma}=0$. 
The dependence of the curvature scalars on the torsion can be made explicit through the relations
\be\begin{split}
&\mathcal{R}=R[g]+2\nabla_{\mu}T^{\mu}-\frac{2}{3}T_{\mu}T^{\mu}+\frac{1}{24}\hat{T}_{\mu}\hat{T}^{\mu}+\frac{1}{2}\tau_{\mu\nu\rho}\tau^{\mu\nu\rho}\\\\&
\Tilde{\mathcal{R}}=-\nabla_{\mu}\hat{T}^{\mu}+\frac{2}{3}\hat{T}_{\mu}T^{\mu}+\frac{1}{2}\epsilon^{\mu\nu\rho\sigma}\tau_{\lambda\mu\nu}\tensor{\tau}{^\lambda_\rho_\sigma},
\end{split}  
\label{METRIC-0}
\ee
where the Ricci scalar $R[g]$ and the covariant derivative are defined in terms of the standard metric Levi-Civita connection.

A general action depending on the curvature scalars can be equivalently expressed in terms of two auxiliary scalars~\cite{Capozziello:2011et} $\chi$ and $\zeta$ as
\be\begin{split}
{\cal{S}}&=\frac{1}{2}\int\,{\rm d}^4x\,\sqrt{-g}\,F({\cal{R}},\tilde{\cal{R}})\,\\
\,\\&
\,=\,\int\,{\rm d}^4x\,\sqrt{-g}\left[\frac{1}{2}F_{\chi}{\cal{R}}+\frac{1}{2}F_{\zeta}\tilde{\cal{R}}-V(\chi,\zeta)\right]
\end{split}
\ee
where $F_{\chi}=\frac{\partial F}{\partial\chi}$ and $F_{\zeta}=\frac{\partial F}{\partial\zeta}$ and
\be V(\chi,\zeta)=\frac{1}{2}\left(\chi F_{\chi}\,+\,\zeta F_{\zeta}\,-F\right)\,.\ee Introducing the metric expressions ({\ref{METRIC-0}}) into the action, we obtain\footnote{The tensorial part $\tau_{\mu\nu\rho}$, appearing in the curvature scalars quadratically upon variation, will yield equations of motion satisfied with $\tau_{\mu\nu\rho}=0$. Therefore, we might as well initially set it to zero.

Note that torsion couples to matter fermions through the currents $J^{\mu}=\bar{\Psi}\gamma^{\mu}\Psi$ and $J_5^{\mu}=\bar{\Psi}\gamma^5\gamma^{\mu}\Psi$. Solving the equations of motion for the torsion vectors and substituting back into the action, we end up with $\mathcal{O}(1/M_P^2)$ effective interactions $(\bar{\Psi}\gamma^{\mu}\Psi)^2$, $(\bar{\Psi}\gamma^{5}\gamma^{\mu}\Psi)^2$ and $(\bar{\Psi}\gamma^{\mu}\Psi)(\bar{\Psi}\gamma^{5}\gamma_{\mu}\Psi)$, which are not expected to play any role during the inflationary phase of the models. 
}
\be\begin{split}
{\cal{S}}&=\int\,{\rm d}^4x\,\sqrt{-g}\Bigg[\frac{1}{2}F_{\chi}R[g]-T\cdot\nabla F_{\chi}\\&+\frac{1}{2}F_{\chi}\left(-\frac{2}{3}T^2+\frac{1}{24}\hat{T}^2\right)+\frac{1}{3}F_{\zeta}T\cdot\hat{T}+\frac{1}{2}\hat{T}\cdot\nabla F_{\zeta}\\&-\frac{1}{2}\left(\chi F_{\chi}+\zeta F_{\zeta}-F\right)\Bigg]
\end{split} 
\ee
Next, varying with respect to the torsion vectors, we obtain their equations of motion, which are algebraic, and substitute the corresponding solutions back into the action.  In addition, we Weyl-rescale the metric as $g_{\mu\nu}\rightarrow\,F_\chi^{-1} g_{\mu\nu}$, going to the Einstein frame, and get (in Planck mass units)
\be {\cal{S}}=\int\,{\rm d}^4x\,\sqrt{-g}\left[\frac{1}{2}R[g]-\frac{3\left(\nabla\left(F_{\zeta}/F_{\chi}\right)\right)^2}{\left(1+4\left(F_{\zeta}/F_{\chi}\right)^2\,\right)}\,-U(\chi,\zeta)\right] \label{Metric-equivalent 1}
\ee
where $U(\chi,\zeta)=V(\chi,\zeta)/F_{\chi}^2$.
Introducing $\omega={F_{\zeta}}/{F_{\chi}}$
we finally get the action in the form
\be {\cal{S}}=\int {\rm d}^4x\sqrt{-g}\left[\frac{1}{2}R[g]-\frac{3(\nabla\omega)^2}{(1+4\omega^2)}-U\right].{\label{E-ACTION}}\ee
It is clear that the gravitational sector, apart from the standard graviton, includes a dynamical pseudoscalar represented by $\omega$. The potential, originally dependent on the two independent scalars $\chi$ and $\zeta$, can be equivalently considered as depending on the dynamical scalar $\omega$ and an additional scalar variable which remains auxiliary, having no kinetic term. The latter can be taken to be
$\chi$. This variable is determined by its equation of motion $\frac{\delta U(\omega,\chi)}{\delta\chi}=0$, the solution of which, substituted back into the action gives its final form 
\be{\cal{S}}=\int\,{\rm d}^4x\,\sqrt{-g}\left[\frac{1}{2}R[g]-\frac{3(\nabla\omega)^2}{(1+4\omega^2)}-U(\omega,\chi(\omega))\right].\ee
For any function $F$ the spectrum of the theory is just the graviton and a pseudoscalar. Different choices of $F$ will result in different forms of the effective potential $U$. 
\section{Models}
\label{Models}
In order to proceed and study the inflationary behavior of the potential $U(\omega)$ we need first to disentangle the scalar variables. Only in the quadratic case $F=\chi+\alpha\chi^2/2+\beta\zeta+\gamma\zeta^2/2+\epsilon\chi\zeta$ does the equation $\delta U/\delta\chi=0$ turn out to be linear in $\chi$ and we can proceed to fully determine $U=(\omega-\beta)^2/4(\gamma+\alpha\beta^2-2\beta\epsilon)$. Nevertheless, since $\chi$ is not associated with any dynamical degree of freedom we could restrict ourselves to the case of an Einstein-Hilbert type of term for the Ricci scalar ${\cal{R}}$ and study a more general dependence of the action on the Holst term. Next, we proceed to consider and analyze two examples of Einstein-Cartan models that go beyond the quadratic dependence of the curvature scalars including cubic terms. The first is a model featuring quadratic and cubic Holst terms, while the second is a model invariant under Weyl transformations with $F\sim \chi^2g(\zeta/\chi)$ for which $g$ is taken to be a cubic function. The aim is to assess the influence of the cubic terms on the the inflationary profile of these models.
\subsection{A cubic-Holst model} Going beyond the quadratic dependence, we consider a cubic action of the form\footnote{Note that in the case that $\tilde{\mathcal{R}}$ is absent, a general metric-affine $F(\mathcal{R})$ model is entirely equivalent to GR without the presence of any dynamical scalar as in the metric case. In the model analyzed in our article we have focused on the simplest choice of a linear Ricci scalar term anticipating that higher powers would not modify this behavior.}
\be \mathcal{S}=M_P^2\int {\rm d}^4x\,\sqrt{-g}\left[\frac{\mathcal{R}}{2}+\frac{\beta \tilde{\mathcal{R}}}{2}+\frac{\tilde{\mathcal{R}}^2}{12M^2}+\frac{\alpha \tilde{\mathcal{R}}^3}{18M^4}\right]
\label{CUBIC-HOLST}
\ee
where we have entered explicitly the Planck scale and have adopted a parametrization analogous to that of the Starobinsky model in terms of two dimensionless parameters $\alpha$ and $\beta$ and a mass scale $M$. Apart from the cubic Holst term $\tilde{\mathcal{R}}^3$ considered, terms ${\cal{R}}^2\tilde{\cal{R}}$ or ${\cal{R}}\tilde{\cal{R}}^2$ are also possible. Nevertheless, the consistency of the scalar representation of the $F(\mathcal{R},\tilde{\mathcal{R})}$ model in terms of the two auxiliary scalars $\chi$ and $\zeta$, apart from the condition $\partial F/\partial\chi>0$, requires that the second derivative matrix $F_{ij}$ of $F(\chi,\zeta)$ should have non-negative eigenvalues, namely 
\be
\det(F_{ij})=F_{\chi\chi}F_{\zeta\zeta}-F^2_{\chi\zeta}\geq\, 0
\ee
If mixing cubic terms are present, such as $\epsilon_1/9M^4\,\chi^2\zeta$, $\epsilon_2/9M^4\,\chi\zeta^2$, the possibility of an instability arises as we see from 
\be
\begin{split}
\det(F_{ij})&=-\left(\frac{2 \epsilon _2 \zeta }{9 M^4}+\frac{2 \epsilon _1 \chi }{9 M^4}\right)^2\\&+\frac{2\epsilon_1\zeta}{9M^4}\left(\frac{2 \alpha  \zeta }{3 M^4}+\frac{2 \epsilon _2 \chi }{9 M^4}+\frac{1}{3 M^2}\right).
\end{split}
\ee
Therefore, we restricted ourselves to the analysis of the presented model that does not include the mixing terms. In terms of the two auxiliaries, Eq. (\ref{CUBIC-HOLST}) corresponds to
\be 
F(\chi,\zeta)=\chi+\beta\zeta+\frac{\zeta^2}{6M^2}+\frac{\alpha\zeta^3}{9M^4}
\ee
The arising dynamical pseudoscalar is
\be
\omega=\frac{F_{\zeta}}{F_{\chi}}=\beta+\frac{\zeta}{3M^2}+\frac{\alpha \zeta^2}{3M^4}
\label{OMEGA-1}
\ee 
Solving Eq. ({\ref{OMEGA-1}}), we obtain
\be 
\zeta=\frac{M^2}{2\alpha}\Big(-1+\sqrt{1+12\alpha(\omega-\beta)}\Big).
\ee
The resulting potential is
\begin{widetext}
\be
\begin{split}
U(\phi)=&\frac{M_P^2M^2}{(12\alpha)^2}\left(-1+\sqrt{1+12\alpha\left(\frac{1}{2}\sinh(-\sqrt{\frac{2}{3}}\phi+\sinh^{-1}(2\beta))-\beta\right)}\right)^2\\&
\Bigg(1+2\sqrt{1+12\alpha\left(\frac{1}{2}\sinh(-\sqrt{\frac{2}{3}}\phi+\sinh^{-1}(2\beta))-\beta\right)}\Bigg)
 \end{split}
 \label{UEXACT}
\ee
\end{widetext}
where $\omega$ has been replaced by a canonical scalar $\phi$, defined by
\be 
\phi\,=\,- M_P\sqrt{\frac{3}{2}}\sinh^{-1}(2\omega)-\sinh^{-1}(2\beta)\,.
\ee 
The positivity of the expression under the root sign would translate into a range of the field $\phi$. Nevertheless, the analysis of the corresponding potential shows that a positive $\alpha$ eliminates the arising inflationary plateau, whereas for $\alpha< 0$ the plateau tends to remain intact. This can be seen from Fig.~\ref{CUBIC-MODEL}. Therefore, restricting ourselves to negative values for $\alpha$ and $\beta$, all positive values of $\phi$ are allowed. The stability condition of our model is given by $F_{\zeta\zeta}>0$, or $F_{\zeta\zeta}=\gamma+ 2 \alpha \zeta$, and substituting the solution of the auxiliary field we obtain $F_{\zeta\zeta}=\sqrt{4\alpha(\omega-\beta)+\gamma^2}>0$. Although a negative contribution from $\alpha$ is required, we have numerically verified the stability criterion in the region of parameter space where we study cosmic inflation.

For small values of $|\alpha|<<1$ we may expand and obtain the approximate expression
\be 
\begin{split}
\frac{16U}{3M^2M_P^2}\approx\left(\sinh(-\sqrt{\frac{2}{3}}\phi+\sinh^{-1}(2\beta))-2\beta\right)^2\\
\,\\
\left[1-\alpha\left(\sinh(-\sqrt{\frac{2}{3}}\phi+\sinh^{-1}(2\beta))-2\beta)\right)\right]
\end{split}\ee
In the next section, based on the exact potential ({\ref{UEXACT}}), we analyze the inflationary behavior of the model and scan for the range of parameters that corresponds to predictions in agreement with current data. Nevertheless, before closing this subsection, it is worth keeping in mind a known fact, namely that the unmodified quadratic model in the limit of large $\beta$ approches the Starobinsky model~\cite{Gialamas:2025thp}, as follows:
\be U_0\approx \frac{3}{4}M_P^2(M\beta)^2\left\{\left(e^{-\sqrt{\frac{2}{3}}\phi}-1\right)^2+\mathcal{O}(1/\beta)\right\}\,.\ee

\subsection{A Weyl-invariant model} Weyl invariance restricts the function $F(\chi,\zeta)$ to be of the form
\be F(\chi,\zeta)=\chi^2g(\zeta/\chi)\ee 
Following the procedure of the previous section, namely, integrating out the torsion, we end up with an equivalent metric action in the Einstein frame ({\ref{E-ACTION}}) in terms of the dynamical pseudoscalar
\be\omega=F_{\zeta}/F_{\chi}=\frac{g'(\zeta/\chi)}{\left[2g(\zeta/\chi)-(\zeta/\chi)g'(\zeta/\chi)\right]}\ee 
The potential is
\be U\,=\,\frac{1}{2}\frac{g(\zeta/\chi)}{\left[2g(\zeta/\chi)-(\zeta/\chi)g'(\zeta/\chi)\right]^2}\ee
Note that, since only the ratio $\zeta/\chi$ appears in $\omega$ and the potential $U$, we could have, equivalently, used Weyl-invariance and gauge fixed the auxiliary scalar $\chi$ to a constant. 

The simplest and known case corresponds to a quadratic action with~\cite{Karananas:2021gco, Gialamas:2024iyu, Karananas:2025fas, Karananas:2025qsm},
\be F(\chi,\zeta)=\frac{\alpha}{2}\chi^2+\frac{\beta}{2}\zeta^2+\frac{\gamma}{2}\chi\zeta.\ee
For this case the potential turns out to be 
\be\begin{split} U(\phi)
\,=\,\frac{1}{4\alpha}+\frac{\alpha}{16|\alpha\beta-\gamma^2|}\left[\sinh\left(\sqrt{\frac{2}{3}}\phi+C\right)-\frac{2\gamma}{\alpha}\right]^2
\end{split}
\ee 
where the canonical scalar $\phi$ is given by \footnote{Setting $\phi\rightarrow -\phi$ and taking the limit of very large $\gamma/\alpha$, for which $\sinh^{-1}(2\gamma/\alpha)\approx\ln(4\gamma/\alpha)$, we get for $\gamma/\alpha\rightarrow \infty$ the following:
$$U(\phi)\,\approx\,\frac{1}{4\alpha}+\frac{1}{4\alpha}\left(e^{-\sqrt{\frac{2}{3}}\phi}-1\right)^2$$
which is just the Starobinsky model with an additional cosmological constant contribution due to the presence of $\chi$.}\be\phi=\sqrt{\frac{3}{2}}\sinh^{-1}(2\omega)-\sinh^{-1}(2\gamma/\alpha).\ee 
 Note the characteristic presence of a cosmological constant. Nevertheless, perturbative and nonperturbative contributions due to quantum effects are expected to arise and modify this term independently of the scalar sector expressed by the rest of the potential. Adopting a phenomenological attitude we assume that this term  acquires its negligible value through the above corrections, while the rest of the potential retains its classical effective form, and we concentrate on the latter.
\begin{figure*}[t!]
  \centering
  \begin{subfigure}[b]{0.49\textwidth}
    \includegraphics[width=\textwidth]{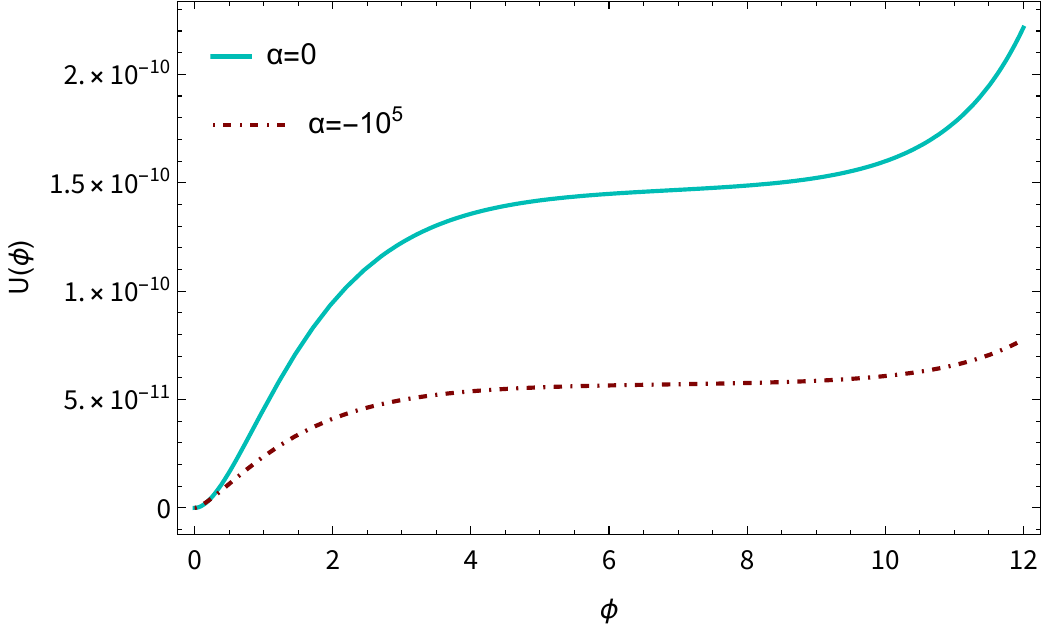} 
    \caption{}
    \label{CUBIC-MODEL}
  \end{subfigure}
  \hfill
  \begin{subfigure}[b]{0.49\textwidth}
    \includegraphics[width=\textwidth]{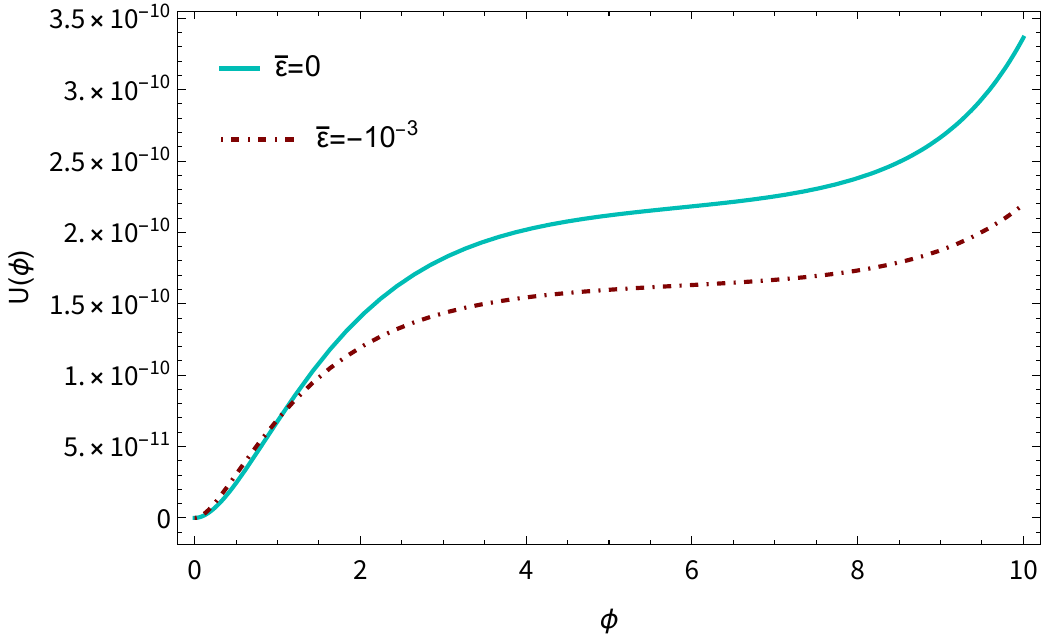}
    \caption{}
    \label{WEYL-MODEL}
  \end{subfigure}
  \caption{(a) Potential for the cubic model, given by Eq.~\ref{UEXACT}. We have used the value $\beta=70$ for the parameter of the linear Holst term, while $\alpha$ is the parameter of the corresponding cubic term. The scale $M^2$ is fixed by the scalar power spectrum amplitude $A_s=2.1\cdot 10^{-9}$. (b) Potential for the Weyl- invariant model, given by Eq.~\ref{WEYL-POTENTIAL}. We have used the parameter values $q=100$ and $q_1=150$, while parameter $\bar{\epsilon}$ scales the cubic correction. The parameter $V_0$ is fixed by the amplitude of the scalar power spectrum  $A_s=2.1\cdot10^{-9}$.  }
  \label{POTENTIALS}
\end{figure*}

Let us consider a deformation of the quadratic case, introducing a parity violating cubic term, of the form\footnote{Note that the validity of the scalar representation rests on the condition $F_{\chi}>0$ and the conditions on the matrix $F_{ij}$, which are not automatic but have to be checked to hold for the range of the parameters considered as well as  for the range of the scalar variable as indeed they do. }
\be g(\zeta/\chi)=\frac{\alpha}{2}+\frac{\beta}{2}(\zeta/\chi)^2+\gamma(\zeta/\chi)+\frac{\epsilon}{3}(\zeta/\chi)^3\ee
Note that $F_\chi=\gamma\zeta-\frac{\epsilon}{3}\frac{\zeta^3}{\chi^2}+\alpha\chi$ and $F_{\zeta}=\beta\zeta+\gamma\chi+\epsilon\frac{\zeta^2}{\chi}$. Gauge fixing as $\chi=1/\alpha$, we obtain in terms of $h=\alpha\zeta$, as follows:
\be \omega=\frac{F_{\zeta}}{F_{\chi}}=\frac{3\left(\gamma+\beta h+\epsilon h^2\right)}{\left(3\alpha+3\gamma h-\epsilon h^3\right)}{\label{OMEG}}\ee
Next, we proceed considering the case that the cubic correction is a small perturbation\footnote{Perturbativity rests on the assumption that $\epsilon h_1<<h_0$ which depends on the parameter range and ultimately corresponds to $\epsilon\Delta U<<U_0$ for the scalar potential $U=U_0+\epsilon \Delta U$. This has been tested numerically to hold comfortably for the range of parameters considered for slow-roll inflation.}
 $|\epsilon|<<\mathcal{O}(\alpha,\,\beta,\gamma)$, setting $h\approx h_0+\epsilon h_1$, and obtain from ({\ref{OMEG}}) to $\mathcal{O}(\epsilon)$,
\be h_0=\frac{\alpha\omega-\gamma}{\beta-\gamma\omega}\,\,\,\,\,{ \textit{and}}\,\,\,\,h_1=\frac{h_0^2(3+\omega h_0)}{3(\gamma\omega-\beta)}\ee
The resulting $\mathcal{O}(\epsilon)$ potential is
\be U\,=\,\frac{1}{4}\frac{(\alpha\omega^2-2\gamma\omega+\beta)}{(\alpha\beta-\gamma^2)}+\frac{\epsilon}{6}\frac{(\gamma-\alpha\omega)^3}{(\alpha\beta-\gamma^2)^2(\beta-\gamma\omega)}\ee
The theory can be expressed in terms of a canonical field $\phi$, defined by
\be \omega\,=\,\frac{1}{2}\sinh(-\sqrt{\frac{2}{3}}\phi+\sinh^{-1}{\frac{2\gamma}{\alpha}})\ee
In our study of inflation we shall consider the case $\gamma>0$ and $\epsilon<0$. In order to facilitate the inflationary analysis we reparametrize the model replacing the parameters $\alpha,\,\beta,\,\gamma,\,\epsilon$ with the set $V_0,\,q,\,q_1,\,\bar{\epsilon}$, defined as\footnote{ The condition $V_0>0$ is equivalent to $q<q_1$ or to $\gamma^2<\alpha\beta$, which is already enforced at the quadratic level as a condition on scalar representation. }
\be 
\begin{split}
&V_0=\frac{\alpha}{16(\alpha\beta-\gamma^2)}>0, \quad q=\frac{2\gamma}{\alpha}, \quad\\\\& q_1=\frac{2\beta}{\gamma},\quad  \bar{\epsilon}=\frac{\alpha^2\epsilon}{3\beta(\alpha\beta-\gamma^2)}
\end{split}
\label{parametrization}
\ee
As in the case of the unperturbed model we focus on the dynamical part of the potential $U(\phi)=1/4\alpha+V(\phi)$, namely
\be 
\begin{split}
  V(\phi)&= V_0 \Bigg[ \left(\sinh\left(-\sqrt{\frac{2}{3}} \phi+ \sinh ^{-1}(q)\right)-q\right)^2\\&+\frac{\bar{\epsilon}  \left(q-\sinh \left(-\sqrt{\frac{2}{3}} \phi+\sinh ^{-1}(q) \right)\right)^3}{1-{q_1^{-1}}{\sinh \left(-\sqrt{\frac{2}{3}} \phi+\sinh ^{-1}(q) \right)}}\Bigg]
\end{split}
\label{WEYL-POTENTIAL}
\ee
Note that in our analysis of slow-roll inflation for the model, carried out in the next section, the chosen range of the parameters $q$, $q_1$ is such that the denominator of Eq.~({\ref{WEYL-POTENTIAL}}) stays nonzero for the full range of $\phi$.

In Fig.~\ref{POTENTIALS}, we show the potentials for the two models for parameter values corresponding to the analyzed inflationary behavior. Note that the presence of a linear Holst term is essential for the formation of the inflationary plateau~\cite{Pradisi:2022nmh, Salvio:2025izr}. 

\section{Inflation}
\label{Inflation}
\begin{figure*}[t!]
  \centering
  \begin{subfigure}[b]{0.49\textwidth}
    \includegraphics[width=\textwidth]{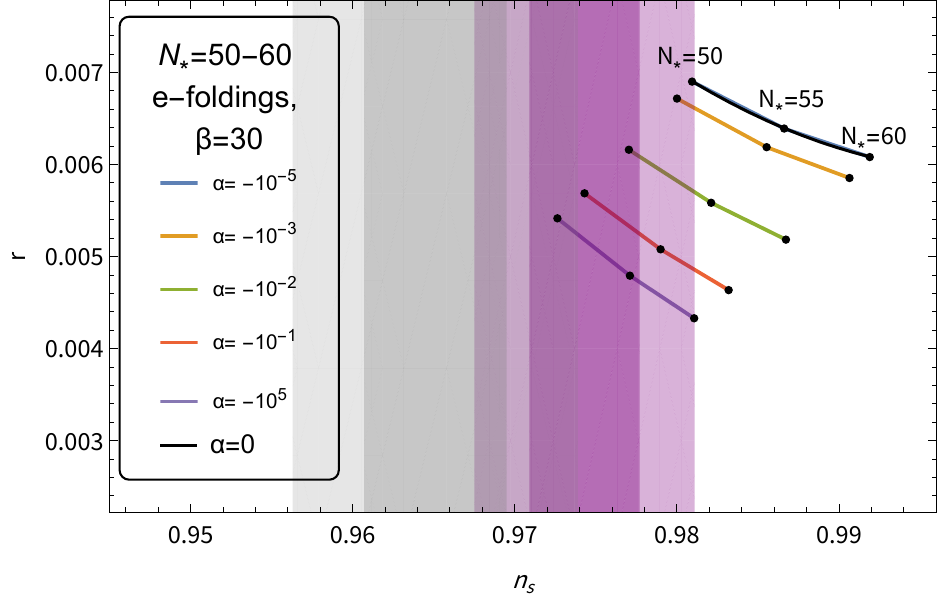} 
    \caption{}
    \label{subfig1-SPECT1}
  \end{subfigure}
  \hfill
  \begin{subfigure}[b]{0.49\textwidth}
    \includegraphics[width=\textwidth]{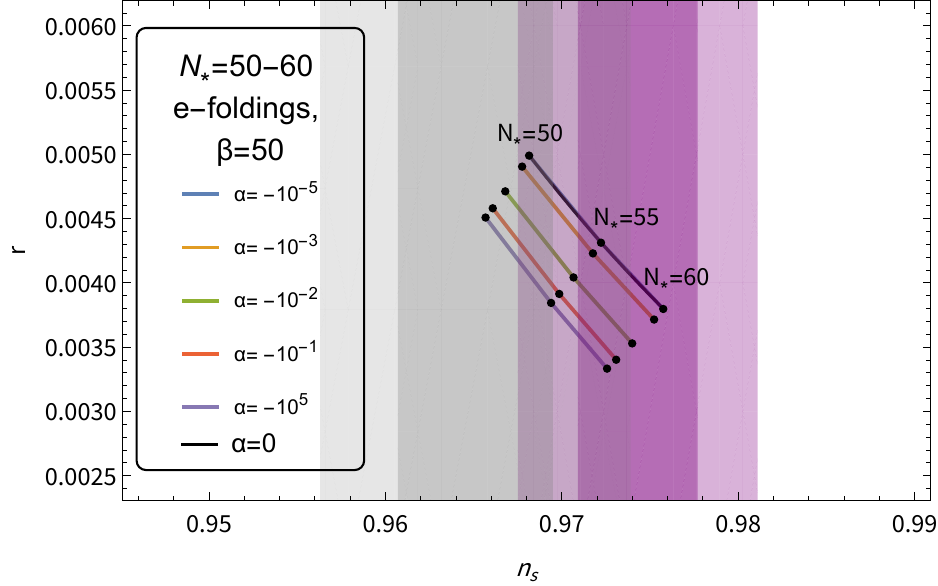}
    \caption{}
    \label{subfig2-SPECT1}
  \end{subfigure}
  \caption{$r$ vs $n_s$ for the cubic-Holst model for two different values of the parameter $\beta$ that parametrizes the linear Holst term, while $\alpha$ corresponds to $\tilde{\mathcal{R}}^3$. The scale $M^2$ is fixed by the observational value of the scalar power spectrum amplitude $A_s=2.1\cdot10^{-9}$. Dark gray and light gray areas denote the $1\sigma$ and $2\sigma$ regions respectively for  the Planck/BK18/BAO dataset~\cite{Planck:2018jri, Planck:2018vyg}, while the dark purple and light purple areas correspond to the Planck/ACT/LB/BK18 dataset~\cite{ACT:2025fju, ACT:2025tim}.}
  \label{SPECT-1}
\end{figure*}
\begin{figure*}[t!]
  \centering
  \begin{subfigure}[b]{0.49\textwidth}
    \includegraphics[width=\textwidth]{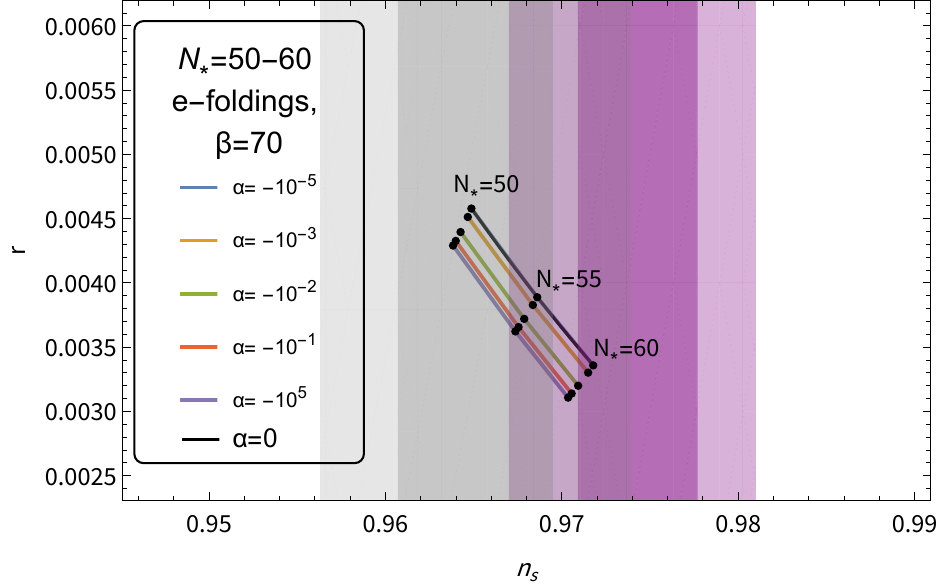} 
    \caption{}
    \label{subfig1-SPECT2}
  \end{subfigure}
  \hfill
  \begin{subfigure}[b]{0.49\textwidth}
    \includegraphics[width=\textwidth]{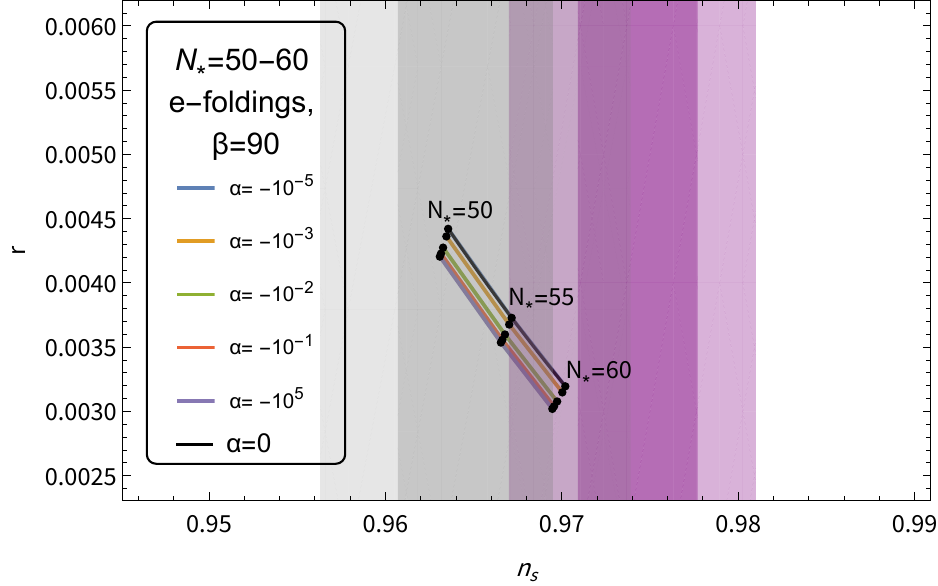}
    \caption{}
    \label{subfig2-SPECT2}
  \end{subfigure}
  \caption{$r$ vs $n_s$ for the cubic-Holst model for two different values of the parameter $\beta$ that parametrizes the linear Holst term, while $\alpha$ corresponds to $\tilde{\mathcal{R}}^3$. The scale $M^2$ is fixed by the observational value of the scalar power spectrum amplitude $A_s=2.1\cdot10^{-9}$. Dark gray and light gray areas denote the $1\sigma$ and $2\sigma$ regions respectively, for the Planck/BK18/BAO dataset~\cite{Planck:2018jri, Planck:2018vyg}, while the dark purple and light purple areas correspond to the Planck/ACT/LB/BK18 dataset~\cite{ACT:2025fju, ACT:2025tim}.}
  \label{SPECT-2}
\end{figure*}
We proceed to analyze the inflationary profile of the single-field models considered in the framework of the slow-roll approximation. We start with the scalar (${\cal{P}}_{\zeta}$) and tensor (${\cal{P}}_T$) power spectra selecting a pivot scale $k_{\star}$ that exited the horizon. 
Their expressions are 
\be
\label{eq:spectra}
\begin{split}
&\mathcal{P}_\zeta (k)=A_s \left(\frac{k}{k_\star} \right)^{n_s -1},\quad{{\rm where}}\quad A_s\simeq\frac{1}{24\pi^2}\frac{{U}(\phi_\star)}{\epsilon_{{U}}(\phi_\star)}\\\\&\quad{{\rm and}}\quad \mathcal{P}_T (k)\simeq\frac{2{U}(\phi_\star)}{3\pi^2} \left(\frac{k}{k_\star} \right)^{n_t},
\end{split}
\ee
with $A_s$ being the amplitude of the power spectrum of scalar perturbations.
The scalar ($n_s$) and tensor ($n_t$) spectral indices given by
\be
\begin{split}
\label{eq:index}
&n_s-1=\frac{{\rm d} \ln \mathcal{P}_\zeta (k) }{{\rm d} \ln k} \simeq -6\epsilon_{{U}} +2\eta_{{U}}  \quad {\rm and}\\\\& \quad n_t= \frac{{\rm d} \ln \mathcal{P}_T (k) }{{\rm d} \ln k}\,,
\end{split}
\ee
\begin{figure*}[t!]
  \centering
  \begin{subfigure}[b]{0.49\textwidth}
    \includegraphics[width=\textwidth]{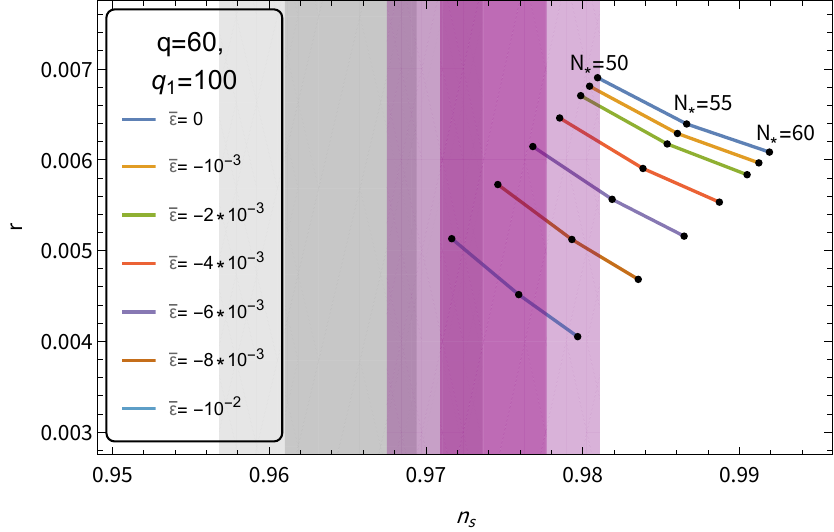} 
    \caption{}
    \label{subfig1-SPECT3}
  \end{subfigure}
  \hfill
  \begin{subfigure}[b]{0.49\textwidth}
    \includegraphics[width=\textwidth]{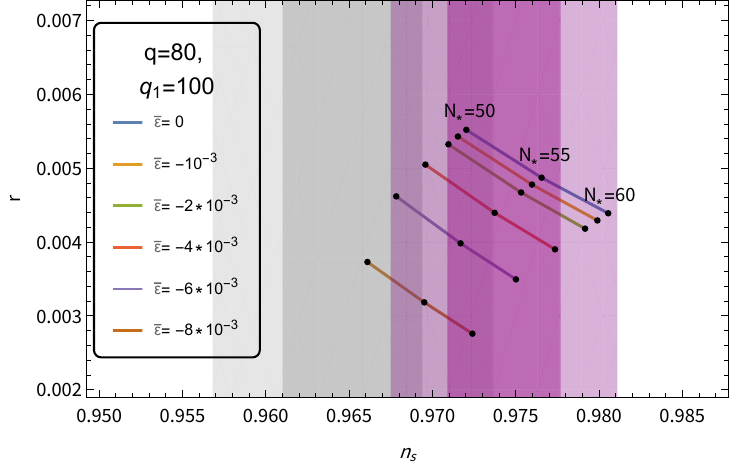}
    \caption{}
    \label{subfig2-SPECT3}
  \end{subfigure}
  \caption{$r$ vs $n_s$ for the Weyl-cubic model for different values of the parameters $q$ and $q_1$, given by Eq.~\ref{parametrization}. The parameter $V_0$ is fixed by the observed value of the scalar power spectrum amplitude $A_s=2.1\cdot 10^{-9}$, while the parameter $\bar{\epsilon}$ scales the cubic term. Dark gray and light gray areas we have denoted the $1\sigma$ and $2\sigma$ regions, respectively, for the Planck/BK18/BAO dataset~\cite{Planck:2018jri, Planck:2018vyg}, while the dark purple and light purple areas correspond to the Planck/ACT/LB/BK18 dataset~\cite{ACT:2025fju, ACT:2025tim}.}
  \label{SPECT-3}
\end{figure*}
\begin{figure*}[t!]
  \centering
  \begin{subfigure}[b]{0.49\textwidth}
    \includegraphics[width=\textwidth]{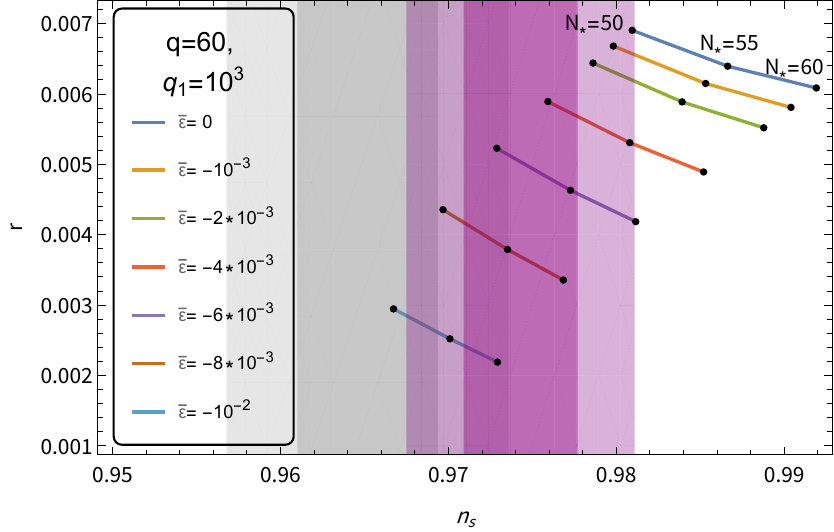} 
    \caption{}
    \label{subfig1-SPECT4}
  \end{subfigure}
  \hfill
  \begin{subfigure}[b]{0.49\textwidth}
    \includegraphics[width=\textwidth]{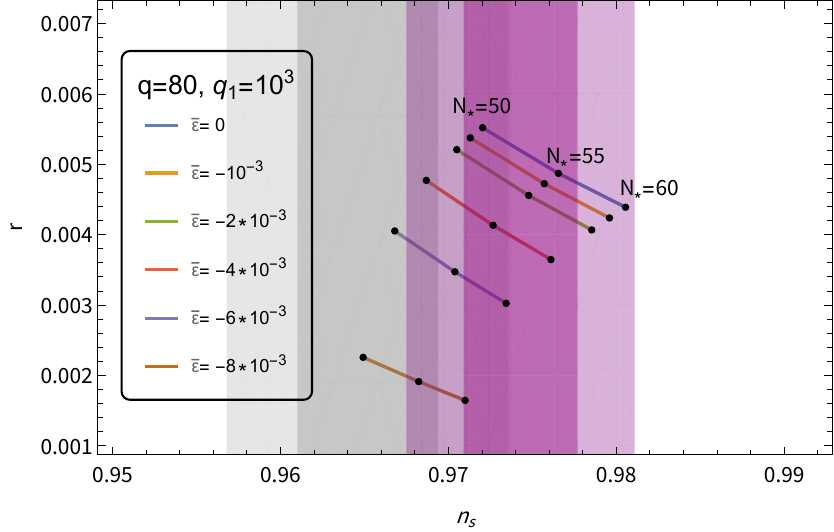}
    \caption{}
    \label{subfig2-SPECT4}
  \end{subfigure}
  \caption{$r$ vs $n_s$ for the Weyl-cubic model for different values of the parameters $q$ and $q_1$, given by Eq.~\ref{parametrization}. The parameter $V_0$ is fixed by the observed value of the scalar power spectrum amplitude $A_s=2.1\cdot 10^{-9}$, while the parameter $\bar{\epsilon}$ scales the cubic term. Dark gray and light gray areas we have denoted the $1\sigma$ and $2\sigma$ regions respectively for  the Planck/BK18/BAO dataset~\cite{Planck:2018jri, Planck:2018vyg}, while the dark purple and light purple areas correspond to the Planck/ACT/LB/BK18 dataset~\cite{ACT:2025fju, ACT:2025tim}.}
  \label{SPECT-4}
\end{figure*}
\begin{figure*}[t!]
  \centering
  \begin{subfigure}[b]{0.49\textwidth}
    \includegraphics[width=\textwidth]{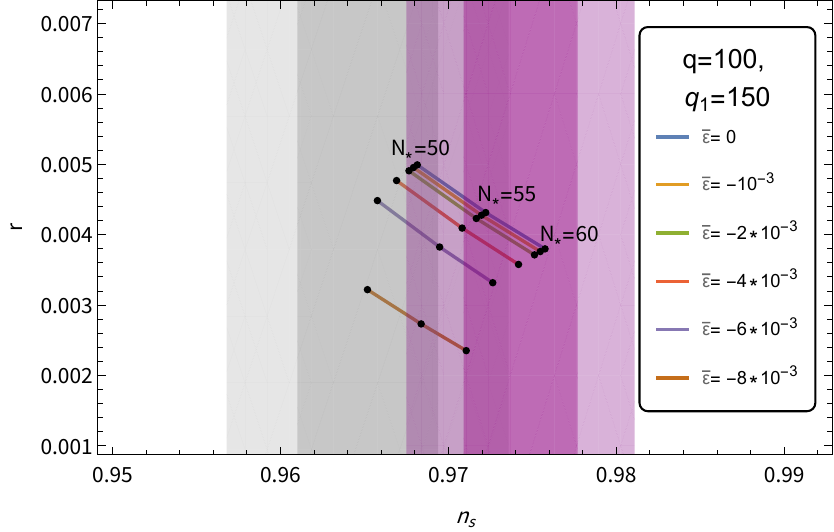} 
    \caption{}
    \label{subfig1-SPECT5}
  \end{subfigure}
  \hfill
  \begin{subfigure}[b]{0.49\textwidth}
    \includegraphics[width=\textwidth]{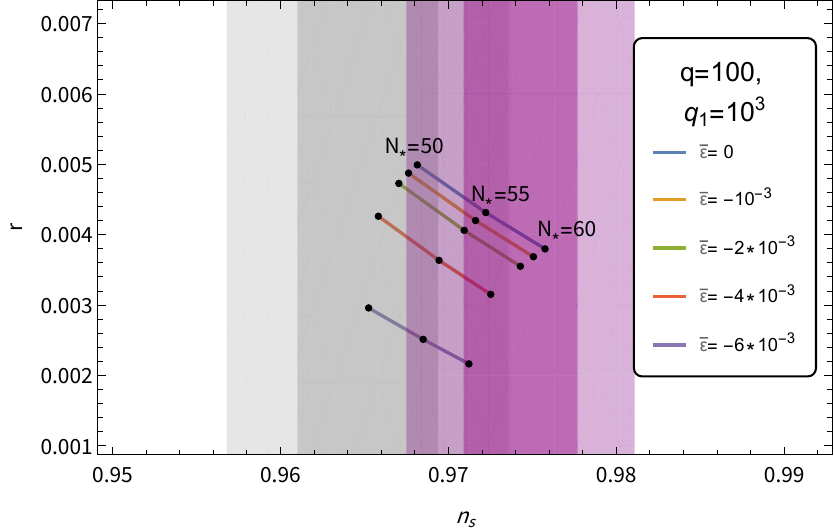}
    \caption{}
    \label{subfig2-SPECT5}
  \end{subfigure}
  \caption{$r$ vs $n_s$ for different choice of Weyl-cubic model parameters. The parameter $V_0$ is fixed by the observed value $A_s=2.1\cdot 10^{-9}$. The parameters $q$ and $q_1$ are given by Eq.~\ref{parametrization} while the parameter $\bar{\epsilon}$ corresponds to the $\tilde{\mathcal{R}}^3/\mathcal{R}$ term. With dark gray and light gray we denote the $1\sigma$ and $2\sigma$ regions respectively for Planck/BK18/BAO dataset~\cite{Planck:2018jri, Planck:2018vyg} while the dark purple and light purple areas corresponds to the Planck/ACT/LB/BK18 dataset~\cite{ACT:2025fju, ACT:2025tim}.}
  \label{SPECT-5}
\end{figure*}
characterize the scale-dependence of the power spectra~\eqref{eq:spectra}. The tensor-to-scalar ratio is defined as\be
\label{eq:ttsr}
    r= \frac{\mathcal{P}_T (k)}{\mathcal{P}_\zeta (k)} \simeq 16\epsilon_{{U}}.
\ee
These equations are expressed in terms of the potential slow-roll parameters 
\be
\label{eq:pslp}
\epsilon_U=\frac{1}{2}\Bigg(\frac{U'(\phi)}{U(\phi)}\Bigg)^2, \quad \eta_U=\frac{U''(\phi)}{U(\phi)}\,.
\ee
The primes denote derivatives with respect the canonical scalar field, while
both slow-roll parameters are assumed to be small $(\ll 1)$ during inflation. 
The duration of inflation is measured by the number of $e$-folds, given by
\be
\label{eq:efolds}
N_\star =\int_{\phi_{\rm end}}^{\phi_\star}\frac{U(\phi)}{U'(\phi)}{\rm d}\phi
\ee
where the end of inflation at $\phi_{\rm end}$ is determined by $ \epsilon_{{U}}(\phi_{\rm end})\,\approx\,1\,$. 

Next, we proceed with the numerical analysis of the observable quantities, namely, the spectral index ($n_s$) and the tensor-to-scalar ratio ($r$). The results are presented in Figs.~\ref{SPECT-1}-\ref{SPECT-5} and Figs.~\ref{SPECT-1} and \ref{SPECT-2} refer to the cubic Holst model and correspond to different values for the parameter $\beta$. This is the dimensionless parameter associated with the linear Holst term in Eq.~\ref{CUBIC-HOLST}. For each value of $\beta$, we have considered selected values of the dimensionless parameter $\alpha$, associated with the cubic term in accordance with the considerations discussed in Sec.~\ref{Models}. We start with the observation that for $\beta= 30$ in the absence of the cubic term ($\alpha=0$) the model fails to produce results compatible with the observational data from Planck/ACT/LB/BK18~\cite{ACT:2025fju, ACT:2025tim} as well as from Planck/BK18/BAO~\cite{Planck:2018jri, Planck:2018vyg}. In this case, agreement is achieved when $|\alpha|\geq 10^{-2}$, with the optimal value being $|\alpha|=10^5$. Focusing in the $\beta$-range from $\mathcal{O}(10)$ to $\mathcal{O}(100)$, we see that as the value of $\beta$ increases the model can remain consistent with the observational data, although the effect of $\alpha$ on the spectral index $n_s$ pushes the model toward the Planck/BK18/BAO region. Nevertheless, for $\beta=70$ and $\beta=90$, the model is in agreement with the Planck/ACT/LB/BK18 data for all values of $\alpha$ as long as $55\leq N_\star\leq60$. In general, the presence of the $\tilde{\mathcal{R}}^3$ term tends to decrease the values of both $n_s$ and $r$. For small values of $\beta$ this helps the model agree with the Planck/ACT/LB/BK18 data while as $\beta$ increases, the model tends to also be consistent with the Planck/BK18/BAO data.

Let us consider next the Weyl invariant, which results from the modification of an existing in the literature quadratic model~\cite{Gialamas:2024iyu, Karananas:2025xcv} by the introduction of a cubic Holst term. The results on the numerical analysis for $n_s$ and $r$  are shown in Figs.~\ref{SPECT-3}-\ref{SPECT-5}. Having reparametrized the model in terms of the parameters $q$ and $q_1$, we first observe that in the case case $q=60$ and $q_1$ is in the range $10^2$ to $10^3$, the model in the absence of the correction (i.e. for $\bar{\epsilon}=0$ ) fails to agree with any of the observational data sets. In contrast, the presence of $\bar{\epsilon}$ succeeds in bringing the model into agreement with the Planck/ACT/LB/BK18 data. We focus on $q=60$ and $q_1=100$ and find that the best agreement with the data is achieved when $|\bar{\epsilon}|\geq 8\cdot 10^{-3}$, whereas for $q=60$ and $q_1=10^3$ we obtain good agreement for $|\bar{\epsilon}|\geq 4\cdot 10^{-3}$. Nevertheless, for a higher value of $q$, e.g. $q=80$ and still $q_1=100$ or $q_1=10^3$, the quadratic model ($\bar{\epsilon}=0$) is consistent with the Planck/ACT/LB/BK18 data, while the presence of the correction tends to push the model toward the region favored by the Planck/BK18/BAO data. The threshold value $|\bar{\epsilon}|=8\cdot 10^{-3}$ appears to mark the point at which the model falls within the $1\sigma$ region. This behavior persists in the case $q=100$ with $q_1=150$ or $q_1=10^3$ , where two limiting values can be identified for which the model agrees with the $1\sigma$ region of the Planck/BK18/BAO data, corresponding to $|\bar{\epsilon}|=6\cdot 10^{-3}$ and $|\epsilon|=4\cdot 10^{-3}$ respectively. In general, the presence of the cubic term in the model leads to a decrease in both $n_s$ and $r$. This helps the model agree with the Planck/ACT/LB/BK18 data for small values of $q$, while for larger values, the presence of this term pushes the model toward the region favored by the Planck/BK18/BAO data.

\section{BRIEF CONCLUSIONS} 
\label{Conclusions}
In the present article we considered the framework of Einstein-Cartan gravity with an action ${\cal{S}}(\mathcal{R},\tilde{\mathcal{R}})$ that depends on both the Ricci scalar curvature and the so-called Holst invariant and focused on models that go beyond a quadratic dependence on them including cubic terms of the latter in the action. We analyzed the slow-roll inflationary behavior of these models in relation to the presently available observational data.  Our numerical analysis showed that in regions of parameter space where the corresponding quadratic model fails due to excessive increase in the predicted spectral index $n_s$, the presence of the cubic term can help to bring the model into agreement with the Planck/ACT/LB/BK18 data. Nevertheless, there are regions where the quadratic models succeed in matching the Planck/ACT/LB/BK18 data, while the inclusion of cubic terms pushes the models toward the region favored by the Planck/BK18/BAO data. Characteristic parameter values for which the {\textit{cubic model}} is in very good agreement with the Planck/ACT/LB/BK18 data lie in the neighborhood of $\beta=30$ and $|\alpha|\gtrsim 10^{-1}$, while for the Weyl invariant model for $q=60$, $q_1=10^2\,\,{\textit{to}}\,\,10^3$ and $|\bar{\epsilon}|\gtrsim 6\cdot10^{-3}$. As a general conclusion we state that going beyond a quadratic dependence on the curvature scalars by the inclusion of cubic terms can have a considerable effect either positive or negative in relation to the agreement with present observational data, depending on parameters.

Summarizing, we may state for the cubic model that acceptable inflation is sustained in the presence of a cubic term as long as the cubic term results in a subdominant correction to the potential in comparison to the contributions of the linear and quadratic terms, while the presence of the linear term is necessary for arising plateau. Note also that in parametric regions where no acceptable inflation arises by the quadratic model, the inclusion of the cubic term can push the corrected model into the acceptable domain. For the Weyl-invariant model, we can state that the general tendency of the cubic term is to decrease the spectral index $n_s$. Therefore in parametric regions where the quadratic model exhibits acceptable inflationary behavior the cubic term tends to push the deformed model outside of the acceptable domain. In contrast, in parametric regions where $n_s$ lies beyond the Planck/BK18/BAO and Planck/ACT/LB/BK18 bounds, the presence of the cubic term helps the model agree with observational data.

\bibliography{references}

\end{document}